\documentclass[aps,prl,reprint,superscriptaddress,showpacs]{revtex4-2}
\usepackage{graphicx}
\usepackage{dcolumn}
\usepackage{bm}
\usepackage{amsmath}
\usepackage{hyperref}
\begin{document}

\title{Analysis of the Gravitational Wave Background Using Gamma-Ray Pulsar Timing Arrays with Next-Generation Detectors}
\author{Zhen Xie}
\affiliation{Deep Space Exploration Laboratory/School of Physical Sciences, University of Science and Technology of China, Hefei 230026, China}
\affiliation{CAS Key Laboratory for Research in Galaxies and Cosmology, Department of Astronomy, School of Physical Sciences,\\
University of Science and Technology of China, Hefei, Anhui 230026, China}
\affiliation{School of Astronomy and Space Science, University of Science and Technology of China, Hefei, Anhui 230026, China}

\author{Zhipeng Zhang}
\affiliation{Deep Space Exploration Laboratory/School of Physical Sciences, University of Science and Technology of China, Hefei 230026, China}
\affiliation{CAS Key Laboratory for Research in Galaxies and Cosmology, Department of Astronomy, School of Physical Sciences,\\
University of Science and Technology of China, Hefei, Anhui 230026, China}
\affiliation{School of Astronomy and Space Science, University of Science and Technology of China, Hefei, Anhui 230026, China}

\author{Jieshuang Wang}
\affiliation{Max-Planck-Institut f\"ur Kernphysik, Saupfercheckweg 1, D-69117 Heidelberg, Germany}

\author{Ruizhi Yang}
\email{yangrz@ustc.edu.cn}
\affiliation{Deep Space Exploration Laboratory/School of Physical Sciences, University of Science and Technology of China, Hefei 230026, China}
\affiliation{CAS Key Laboratory for Research in Galaxies and Cosmology, Department of Astronomy, School of Physical Sciences,\\
University of Science and Technology of China, Hefei, Anhui 230026, China}
\affiliation{School of Astronomy and Space Science, University of Science and Technology of China, Hefei, Anhui 230026, China}

\begin{abstract}

In this work, we investigate the potential of gamma-ray pulsar time array (PTA) on gravitational waves background (GWB) using future gamma-ray detectors with larger effective areas. We consider both spaceborne detectors and ground-based imaging air Cherenkov telescope arrays (IACTs).  We simulated the detected photons from pulsars using the response of hypothetical detectors taking into account the backgrounds and analyzed the sensitivities.  Our results showed that thanks to the higher statistics of IACTs, the PTA using IACTs can improve significantly the performance compared with the PTA using Fermi-LAT data.

\end{abstract}
\maketitle
\section{Introduction}
Pulsars are ideal cosmic laboratories for their excellent periodicity. 
The pulsar timing array (PTA) is the only method so far to detect the low-frequency gravitational waves (GWs) in nHz
\cite{1979ApJ...234.1100D}.  
The GWs can be detected using ensembles of millisecond pulsars (MSPs) known as pulsar timing arrays (PTAs). PTAs monitor the arrival times of steady pulses from each pulsar, which are affected by spacetime perturbations and may arrive earlier or later than expected.
For observations taken on Earth, the low-frequency GWs are expected to produce a signature quadrupolar pattern of the TOAs of the photons that come from the pulsar, known as the Hellings-Downs correlation\cite{1983ApJ...265L..39H}.

Low-frequency GWs have many origins, and they can provide a wealth of information about the universe. Supermassive black hole (SMBH) binaries are expected to emit GWs, and the superposition of GWs from many SMBH binaries throughout the universe is predicted to build up a GW background (GWB). GWs from inflation would help describe the universe at its earliest moments\cite{PhysRevD.50.1157} and are also an important way to test cosmology theories. Cosmic strings are theorized topological defects produced by phase transitions in the early universe, vibrating and losing energy via gravitational wave emission over the history of the universe \cite{GASPERINI1993317}.
If cosmic strings exist, they will create a stochastic GWB, and the observation of such kind of GWB would bring confirmation of physics beyond the Standard Model \cite{PhysRevD.86.023503}. As mentioned above, since many processes can produce GW signals, the information derived from stochastic GWB would provide significant information about astrophysical processes over the history of the universe\cite{Christensen_2019}.

Recently, the Fermi-LAT Collaboration has performed for the first time the study of gravitational wave background using PTA observed in gamma-ray band\cite{fermi2022gamma}, which demonstrates the great potential to study the GWB. Gamma PTA has many advantages compared with the traditional ratio PTAs. For example, a main noise source for radio PTAs is the effect of radio propagation through plasma, including the solar wind and the ionized interstellar medium (IISM). These effects are time-dependent and introduce noises similar to the GW signals.  On the other hand, the effects of the IISM and solar wind can be ignored for gamma-ray photons. In this regard,  gamma PTA has smaller noise and much easier data analysis. 

But gamma PTA also suffers from poor angular resolution and limited exposure of the current instrument.
In this letter, we investigated the potential improvement of gamma PTA\footnote{https://zenodo.org/record/6374291\#.YzVcbC-KFpR} by future detectors. We considered two types of instruments. One is future spaceborne telescopes (FSTs) like Fermi-LAT with a larger effective area; and the other is Image Air Cherenkov Telescopes (IACTs), these ground-based telescope has a much larger effective area with high time accuracy.

Our work follows this structure. We described the method we used to simulate the observation of pulsars using the hypothetical instruments in session 2, we analyzed the simulated data and investigated the sensitivities of gamma PTA with future instruments in session 3, and the last session is the conclusion.

\section{simulated data based on future detectors}

In Fermi-LAT gamma PTA, pulsar PSR J1231-1411 gave the best constraint of the photon-by-photon method, so we used this object in the following simulation as an example.

\begin{figure*}[]
		\centering
        \includegraphics[width=0.95\textwidth]{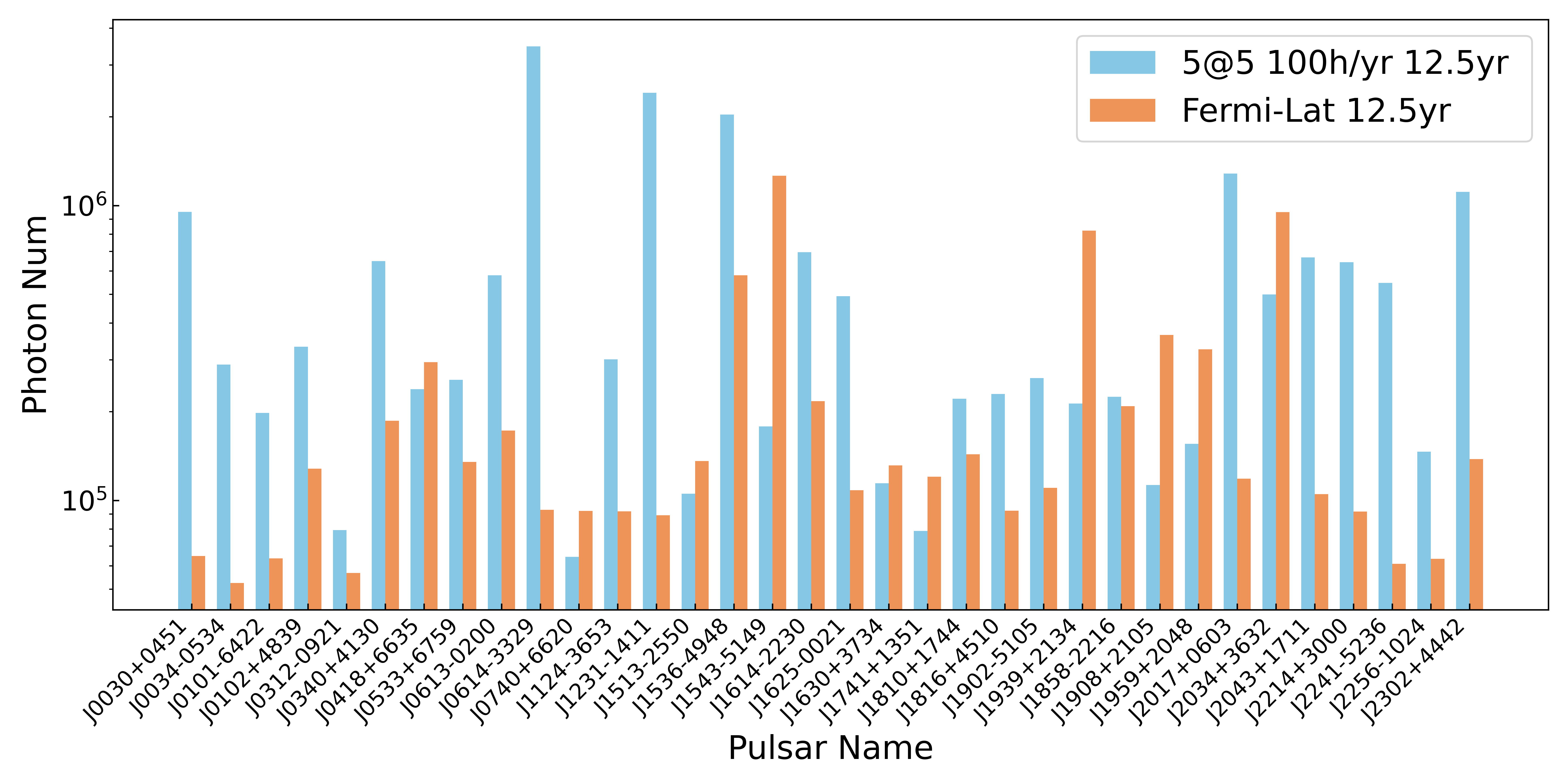} 
        \caption{The number of effective photons of 35 Fermi-LAT pulsars\cite{fermi2022gamma} measured in 12.5 years, compared with the expected number of photons by 5@5 observation 100h/yr in 12.5 years, at the energy from 1 to 10~Gev. }
	\label{Fig:55cpfermi}
\end{figure*}

In the simulation, two different types of detectors are considered. Firstly we consider FSTs similar to Fermi-LAT but with 10 times more effective area. Another type is low threshold IACTs. In this work, we adopt 5@5 as an example of such kind of detector. 
5@5 is a large ground-based Cherenkov telescope array planned for the mountains of the Atacama Desert in northern Chile. Due to its low energy threshold, it shows great potential for pulsar research. 
In this paper, we used the response of 5@5 to perform the simulation. In the analysis we didn't consider the true geometrical location of the arrays, instead, we just assumed a 100-hour exposure of the pulsar with the fiducial telescope response. We admit that the true instrument response will depend on the site location as well as the source declination, but for a single pulsar, it is easy to find 100 hours of observation times every year with reasonable declination. Thus in the work, we used a uniform instrumental response for IACTs for simplicity.
The telescope's effective area can be described as \citet{aharonian2015} calculated:
\begin{equation}
    A_{eff} = 8.5E^{5.2}[1+(E/5~GeV)^{4.7}]^{-1}~\rm m^2,
    \label{eq:spec}
\end{equation}
and the point speared function (PSF) of 5@5 can be described as:
\begin{equation}
    \phi = 0.8(E/1~GeV)^{-0.4}~{\rm degree},
\end{equation}
 by integrating with the spectrum of the pulsar, we can derive the expected photon number of the IACTs. Fig.~\ref{Fig:55cpfermi} shows the result of the photon number by Fermi-LAT and 5@5 which makes an observation for 100~h per year in 12.5 years. We found that the ground-based telescope has a good performance in collecting photons, due to their large effective area. For J1231-1411, we made a conservative estimate to observe it 100h per year, the photon number IACT can collect is 30 times more than that from Fermi-LAT in the same time span. We note that a significant disadvantage of IACTs is the much smaller FOV and lower duty cycles. Fermi LAT results showed that the combined likelihood of more than 20 pulsars can further improve the sensitivity by a factor of two. In this regard, IACTs cannot compete because of the limited sky coverage. But thanks to the advantage of photo sensors, the next generation IACTs can also operate on the night with moon \cite{aharonian2021construction}, thus the observation time every year can be increased to nearly 2000 hours. Thus it would be easy to observe more than 10 pulsars every year with an exposure of about 100 hours each, which will also allow us to perform the joint likelihood analysis.  

For Fermi-LAT, the gamma-ray data are recorded in terms of energy $E_i$, spatial position ${\bf r}_i$, and arrival time $t_i$ for the $i$-th photon. So in simulations for photons detected by hypothetical detectors, we also sample these quantities.  The energy for photon from a pulsar can be described by the parameterized function PLSuperExpCutoff4 used by Fermi-LAT\cite{abdollahi2022incremental}:
\begin{equation}
    \frac{dN}{dE} = K(\frac{E}{E_0})^{\frac{d}{b}-\Gamma_s}exp[\frac{d}{b^2}(1-(\frac{E}{E_0})^b)]~~~~~~~(b~ln\frac{E}{E_0}>10^{-2})\label{eq:energy}\,,
\end{equation}
each parameter can be queried in the catalog provided by Fermi-LAT. We first sample the energy of the photons by using this distribution. 
For the spatial position, we chose a circle of $3^\circ$ radius around the pulsar as Fermi-LAT PTA and then sampled the position of the detected photon by taking into account the point spread function (PSF) of the detector, as well as the flux from both pulsar and a flat background. Note that the PSF is always energy-dependent. 

 Due to the high, sometimes even dominating backgrounds in gamma-ray astronomy, it is always difficult to recognize whether the photon comes from the pulsar itself or from backgrounds. The background in Fermi LAT (and other space-borne detectors) is mainly the diffuse Galactic gamma-ray background (DGE). In Fermi LAT it is described in the standard background file \textit{gll\_iem\_v07.fits} \cite{acero2016development}. It is taken into account in the data analysis in Fermi PTA to calculate the \textit{weight} of photons. As in gamma PTA, \textit{weight} is given to each photon to show the possibility of whether the photon comes from a pulsar or not. 
 
In IACTs, however, in addition to the DGE, there are unavoidable contaminations from cosmic ray (CR) proton and electrons which are also observed by IACTs. In the energy range we are interested in this work ($1 - 10~\rm GeV$), the CR electrons cannot be detected by IACTs due to the geomagnetic cutoff effects. As calculated in \cite{cheback}, the background from CR protons can also be neglected in this energy range due to a much lower trigger rate at low energy. In this case, the dominating background in IACTs would also be DGE, and the analysis for IACTs would be identical to that of Fermi-LAT and FSTs. 

However, we cannot exclude the possibility that IACT could induce further CR backgrounds due to different configurations to that used in \cite{cheback}.  As a conservative check, in this work we estimated the CR proton background based on the results in \citet{aharonian2015}, the background for 1 - 10~\rm GeV gamma-rays mainly comes from the protons with energy 10 - 100~\rm GeV, considering also a gamma/p separation power of about 1/10, the flux of background from CR protons can be written as $F_{\rm bkg}=2\times 10^{-7}~\rm(E/1GeV)^{-2.7}MeV^{-1}sr^{-1}cm^{-2}s^{-1}$,  which is at least one order of magnitude larger than the DGE in the plane at the same energy range. As a result, we consider only the background induced by CR protons in the calculation for IACTs. We also assume it is uniformly distributed spatially due to the homogeneity of CR proton arriving directions.  In addition to the primary electron and CRs, the secondary electrons produced in the primary CR interaction with the atmosphere can be another background. But these secondary electrons should be part of the hadronic shower induced by primary CR protons, which is already included in the proton background and gamma/p separation procedure discussed above.

The arrival time of each photon can be translated into the phase by the \textit{PINT} software\cite{luo2021pint}, by accumulation, we can get the pulse profile. And the profile can be described by the superposition of several Gaussian distributions, which is called the template function. In our simulation, we used the profile folded by Fermi-LAT Observation data in 12.5 years. The sampling of the arrival time of a photon consists of two parts, integer multiples of the period of the pulsar and the phase (time) conforming to the pulsar's pulse profile, which is described of template functions for PSR J1231-1411 derived in Fermi PTAs \citep{fermi2022gamma}.

The last step of simulating is to calculate the \textit{weight} of each photon. We calculated the predicted photon flux from the  pulsar by convolving the flux of the pulsar with the PSF at each position, as well as the flux from the background. We calculated the \textit{weight} of each photon by dividing the photon flux from the pulsar by the total photon flux (pulsar plus background) at each position. 

Through the above steps, we simulated the energy, time(phase), position, and the \textit{weight} information of each incident photon, we used them in the analysis using gamma PTA pipelines. 

\section{gamma PTA Data analysis}
 The  log-likelihood function of a single pulsar is given by unbinned (photon-by-photon) method\cite{fermi2022gamma}:
\begin{equation}
\begin{aligned}
	\log \mathcal{L} &= \sum_i \log \left[ w_i f(\phi_i) + (1 - w_i) \right] - \\ 
    &0.5\beta^T C^{-1}_{\rm tn}\beta - \frac{1}{2}\log(|C_{\rm tn}|), \label{Eq:loglik}
\end{aligned}
\end{equation}
here,$\phi$ is the phase of an individual pulsar, and $f(\phi)$ is the profile of $\phi$, which is defined by a sum over one (or many) Gaussian distributions $g(\phi; \mu, \sigma)$ with the mean $\mu$ and the variance $\sigma$, and each photon is assigned a \textit{weight} which characterizes its probability of originating from the pulsar or background as described earlier. The second part represents a Gaussian noise process of Fourier amplitudes ${\bm \beta}$:
\begin{align}
		\mathcal{L}_{\rm tn} \propto \dfrac{1}{\sqrt{|C_{\rm tn}|}}\exp\left(-\dfrac{1}{2}\beta^T C^{-1}_{\rm tn}\beta \right).
	\end{align}

To compare these results with radio PTA, we assumed that both IACTs and FSTs will start observation in 2035 when such large instruments are likely to be put into operation. Then we calculated the sensitivities with observation duration.  We considered the constraints for the single source PSR J1231-1411, which gave the best constraints for Fermi-LAT. For IACTs, we calculated both the sensitivities with and without the hypothetical CR background, assumed an effective exposure time of 100 hours every year. We found the IACTs we considered 
\begin{figure*}[]
		\centering
        \includegraphics[width=0.95\textwidth]{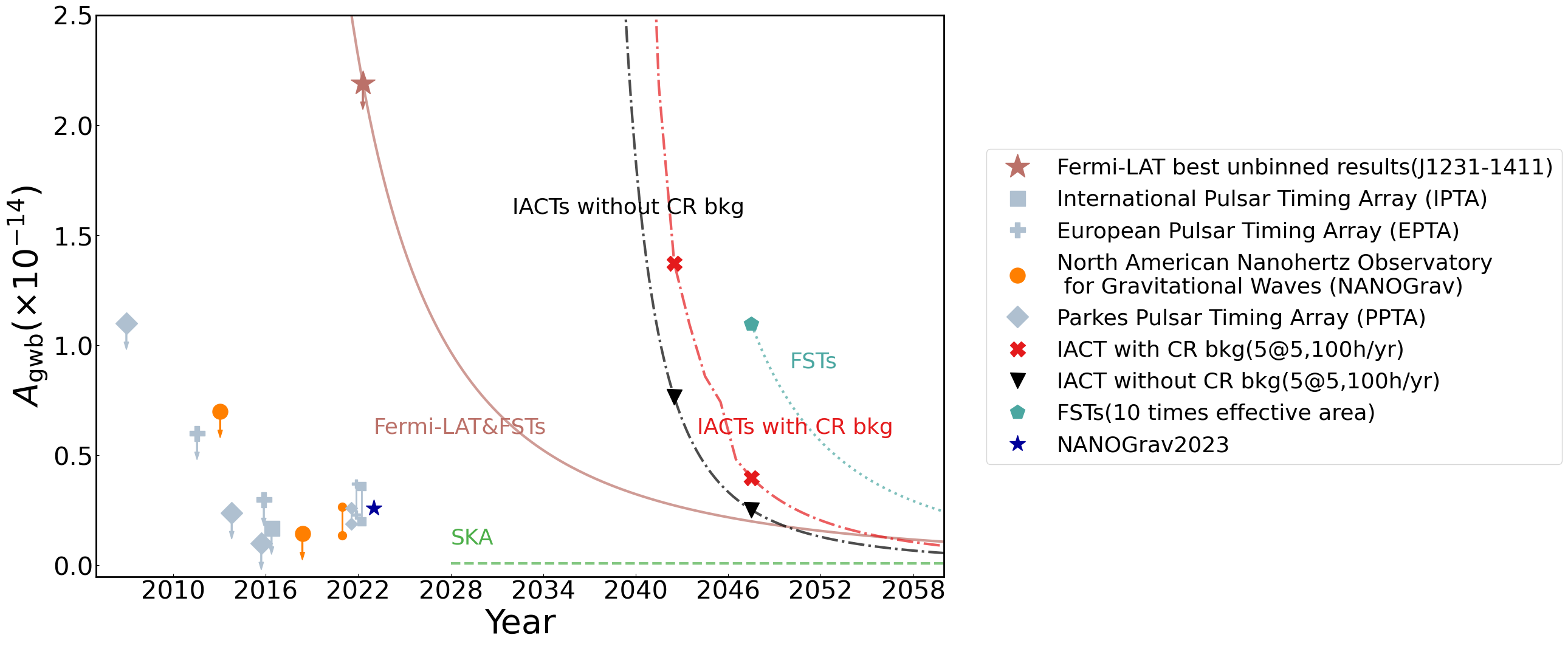} 
        \caption{Constraints on the GWB from radio and gamma-ray PTAs, the radio PTA data is from Fermi-LAT\cite{fermi2022gamma}. Assuming that both IACTs and FSTs start observation in 2035 {\bf(note that the  data points before 2045 are above the $A_{gwb}$ range shown in this figure due to the steep rise of the sensitivity curve)}, the points in the right half show the results for about 7.5 and 12.5-year observations of J1231-1411. The solid line shows the Fermi-LAT result, in which the sensitivity is proportional to  $t_{obs}^{-13/6}$. The dot-dash line shows the results of IACTs with and without background as Fig.~\ref{Fig:modts2}. The green line shows the level at which SKA can be reached when it goes into operation in 2028. The orange star is the NANOGrav 15-year Data Set result.}
	\label{Fig:predict}
\end{figure*}
here have a sensitivity significantly better than FSTs, even though we have assumed that the FSTs have a 10 times larger effective area than Fermi-LAT, which is nearly unrealistic.  The Gamma PTA with IACTs can surpass the Fermi LAT sensitivities within a decade or so after the operation.
We also compared these results with the recent NANOGrav 15-year Data Set result, which gave evidence of GWB with the amplitude of $2.6\times10^{-15}$ at a reference frequency of 1 yr$^{-1}$\cite{2023ApJ...951L...8A}. The Square Kilometre Array (SKA) can greatly enhance pulsar timing precision by its unprecedented collecting area and bandwidth, and the expected levels to be reached by the SKA is about $10^{-16}-10^{-17}$ at a reference frequency of 1 yr$^{-1}$
\cite{2015aska.confE..37J}. These results are shown in Fig.~\ref{Fig:predict}.

\begin{figure}[h]
		\centering
        \includegraphics[width=0.45\textwidth]{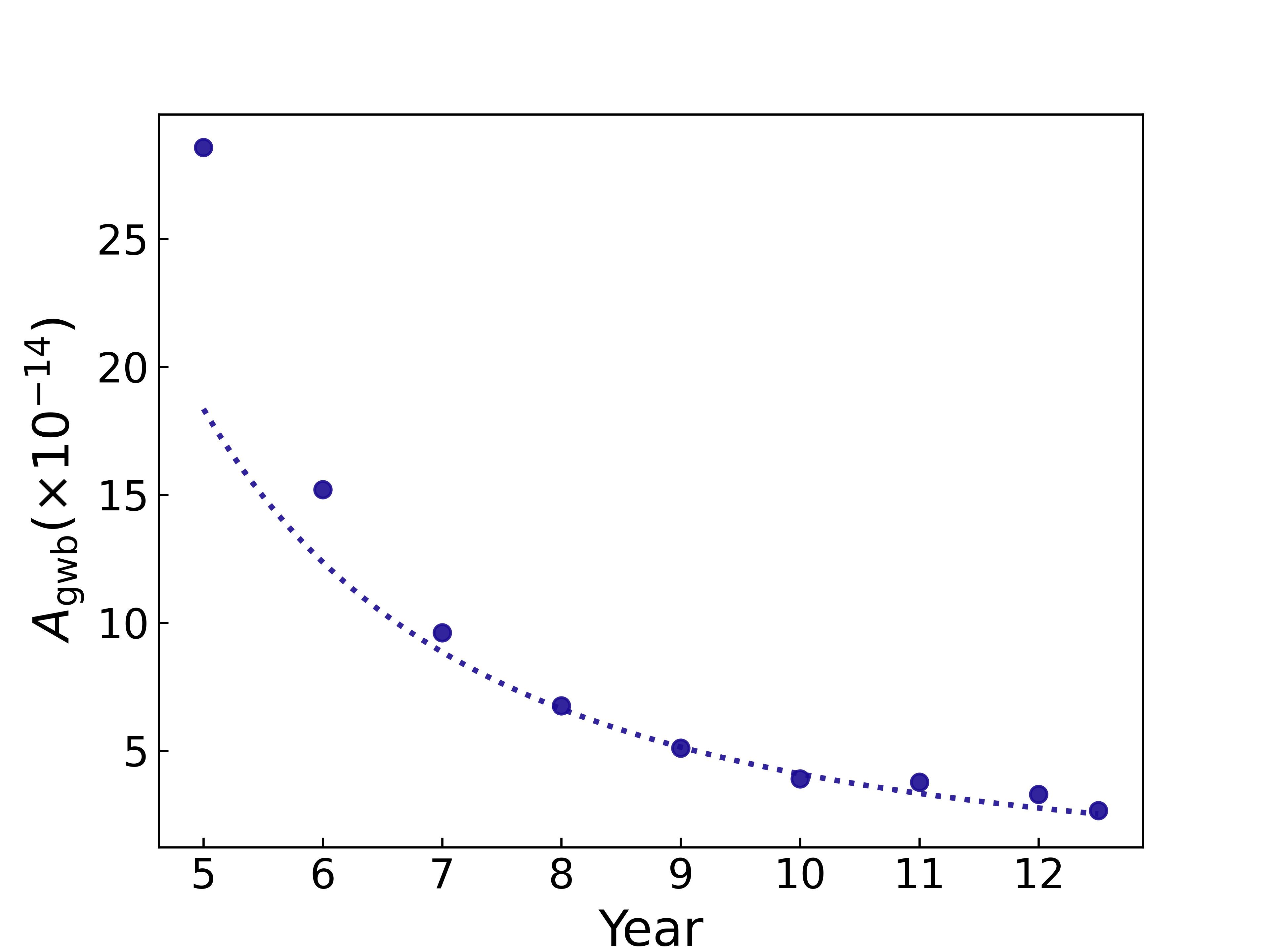} 
        \caption{Changing in $A_{gwb}$ limit for J1231-1411 with increased observation time using Fermi-LAT data. The dashed line represents the relationship that $A_{gwb}$ is proportional to $t_{obs}^{-13/6}$.}
	\label{Fig:realts}
\end{figure}

\begin{figure}[h]
		\centering
        \includegraphics[width=0.45\textwidth]{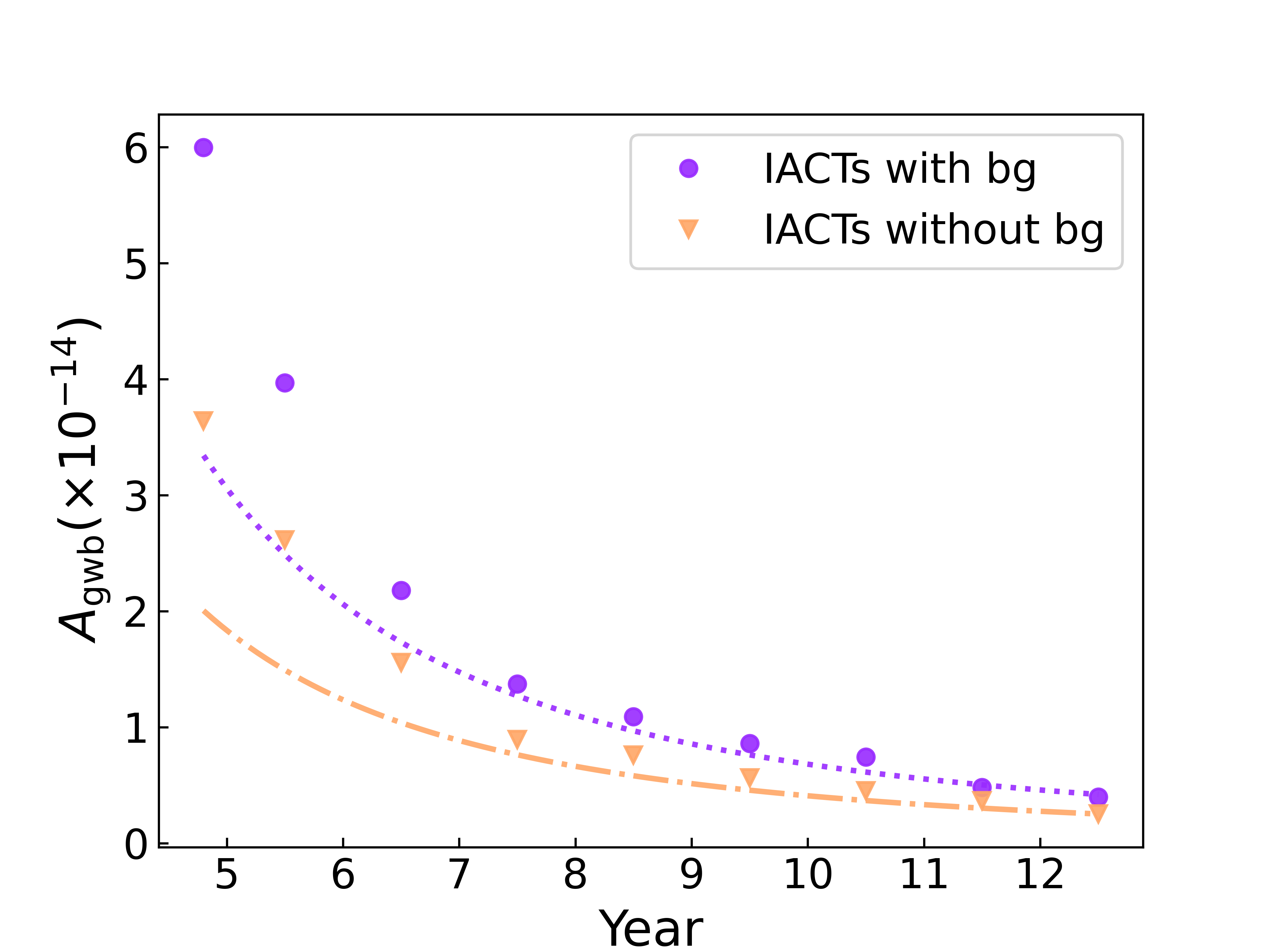} 
        \caption{Changing in $A_{gwb}$ limit for J1231-1411 by simulated data of the IACTs with and without backgrounds. The dashed line represents the relationship that $A_{gwb}$ is proportional to $t_{obs}^{-13/6}$.}
	\label{Fig:modts2}
    \end{figure}

\begin{figure}[h]
		\centering
        \includegraphics[width=0.45\textwidth]{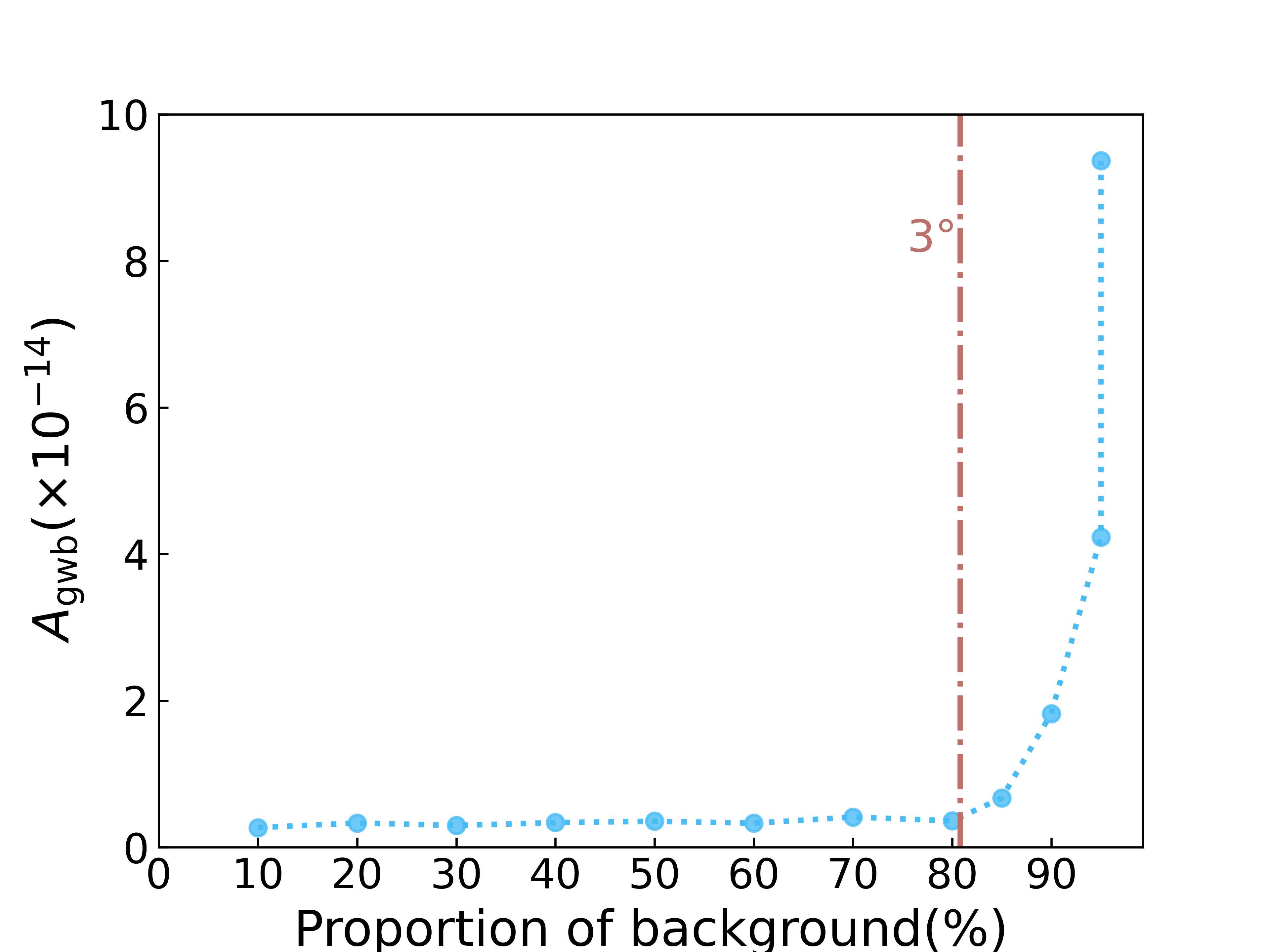} 
        \caption{The influence of $A_{gwb}$ by different background levels. The sensitivity has been induced sharply in high background ratios, which may be due to the profile of the pulsar being affected. The $3^\circ$ is the radius of the exposure window used in Fermi-LAT gamma PTA and this work.}
	\label{Fig:bgg}
    \end{figure}

 For ideal PTA,the signal-to-noise ratio grows proportionally to $A_{gwb}^2\times t^\Gamma _{obs}$\cite{pol2021astrophysics}. So as $\Gamma = 13/3$ for SMBH generated GWB\cite{2004ApJ...611..623S}, the relation of $A_{gwb}$ with the observation time length $t_{obs}$ will be $A_{gwb}\propto t_{obs}^{-13/6}$, here the dimensionless strain amplitude $A_{gwb}$ incorporates the growth, masses, and merger rates of SMBHs, and the $\Gamma$ is the spectral index of spectrum of GWBs power spectral densities.  We calculated the Fermi LAT upper limit on $A_{gwb}$ with different $t_{obs}$ using the real Fermi-LAT data and the results are shown in Fig.~\ref{Fig:realts}. 
We also calculated the sensitivity with time using IACTs, assuming an exposure of 100 hours per year. The results are shown in Fig.~\ref{Fig:modts2}. We can see that the sensitivity gradually gets closer to the expectation for both Fermi-LAT and IACTs with the increase of observation time. 
In order to consider the level of how the background of IACTs influences the sensitivity of GWB analysis, we also simulated data with different background levels, as Fig.~\ref{Fig:bgg} shows. From our calculated results, we found that the influence is relatively small when the background contributes less than 80\% of the total photons since the photons that come from the background have lower \textit{weight}, which seldom affects the pulsars profile. While at higher rates, it weakens the sensitivity sharply, this may be due to the profile of the pulsar being broken by the background. In very high background ratios($>95\%$), even fitting the profile of the pulsar is failed. So it's still necessary to lower the background photon's effect in ground-based observation. A possible way is using a small exposure window near the center of the pulsar. In our current work, we used $3^\circ$ region, a smaller exposure window can improve the performance of sensitivity. Considering the PSF of about $1^\circ$ for IACTs in this energy range, such improvement is feasible.

\section{Discussion} 
In this paper, we extended the gamma PTA analytical method of GWB used by Fermi-LAT to simulate future gamma-ray detectors' capability on gamma PTA.  Both IACTs and FSTs can lift the statistics significantly. IACTs would potentially induce extra CR backgrounds, which could limit the sensitivity to GWB.  We took the extra background into account and found that, in our conservative estimation of CR backgrounds, the IACTs  still gave a much  better sensitivity due to their overwhelming effective area. 

Meanwhile, the sensitivity of gamma PTA is still limited, and there's still a gap with the result of radio PTA. This is not only due to the limitation of existing instruments but also on account of the short time length of gamma PTA observations. The sensitivity of gamma PTA is hard to compare with radio PTA in the short term. But as we had discussed in this letter, gamma PTA shows great potential to match radio PTA in a decade, especially with future detectors. Beyond that, due to the much easier data reduction procedure and less impact from ISM plasma for gamma PTA, we believe that the cross-check from multi-wavelength observation is also necessary and important to limit the GWB and other physical progress.

Looking ahead, large gamma-ray instruments have been planned or are already under construction, such as  VLAST\cite{2022AcASn..63...27F} and Chrenkov Telescope Array (CTA) \cite{hofmann2023cherenkov}. There is also a plan to build IACTs on the site of the Large High Altitude Air Shower Observatory (LHAASO) \cite{zhen2019introduction}, \cite{lacticrc}. But for low threshold IACTs the LHAASO site may be not good enough because of the limited weather conditions, other better sites for optical astronomy, such as Lenghu \cite{deng2021lenghu} are more suitable for such an instrument.  The gamma PTA is a supplementation and cross-checking tool for radio PTA.  With the continued development of new detection tools, we expect further progress in understanding these elusive phenomena. 

\appendix

\bibliography{main.bib}

\begin{thebibliography}{23}%
\makeatletter
\providecommand \@ifxundefined [1]{%
 \@ifx{#1\undefined}
}%
\providecommand \@ifnum [1]{%
 \ifnum #1\expandafter \@firstoftwo
 \else \expandafter \@secondoftwo
 \fi
}%
\providecommand \@ifx [1]{%
 \ifx #1\expandafter \@firstoftwo
 \else \expandafter \@secondoftwo
 \fi
}%
\providecommand \natexlab [1]{#1}%
\providecommand \enquote  [1]{``#1''}%
\providecommand \bibnamefont  [1]{#1}%
\providecommand \bibfnamefont [1]{#1}%
\providecommand \citenamefont [1]{#1}%
\providecommand \href@noop [0]{\@secondoftwo}%
\providecommand \href [0]{\begingroup \@sanitize@url \@href}%
\providecommand \@href[1]{\@@startlink{#1}\@@href}%
\providecommand \@@href[1]{\endgroup#1\@@endlink}%
\providecommand \@sanitize@url [0]{\catcode `\\12\catcode `\$12\catcode
  `\&12\catcode `\#12\catcode `\^12\catcode `\_12\catcode `\%12\relax}%
\providecommand \@@startlink[1]{}%
\providecommand \@@endlink[0]{}%
\providecommand \url  [0]{\begingroup\@sanitize@url \@url }%
\providecommand \@url [1]{\endgroup\@href {#1}{\urlprefix }}%
\providecommand \urlprefix  [0]{URL }%
\providecommand \Eprint [0]{\href }%
\providecommand \doibase [0]{https://doi.org/}%
\providecommand \selectlanguage [0]{\@gobble}%
\providecommand \bibinfo  [0]{\@secondoftwo}%
\providecommand \bibfield  [0]{\@secondoftwo}%
\providecommand \translation [1]{[#1]}%
\providecommand \BibitemOpen [0]{}%
\providecommand \bibitemStop [0]{}%
\providecommand \bibitemNoStop [0]{.\EOS\space}%
\providecommand \EOS [0]{\spacefactor3000\relax}%
\providecommand \BibitemShut  [1]{\csname bibitem#1\endcsname}%
\let\auto@bib@innerbib\@empty
\bibitem [{\citenamefont {{Detweiler}}(1979)}]{1979ApJ...234.1100D}%
  \BibitemOpen
  \bibfield  {author} {\bibinfo {author} {\bibfnamefont {S.}~\bibnamefont
  {{Detweiler}}},\ }\bibfield  {title} {\bibinfo {title} {{Pulsar timing
  measurements and the search for gravitational waves}},\ }\href
  {https://doi.org/10.1086/157593} {\bibfield  {journal} {\bibinfo  {journal}
  {\apj}\ }\textbf {\bibinfo {volume} {234}},\ \bibinfo {pages} {1100}
  (\bibinfo {year} {1979})}\BibitemShut {NoStop}%
\bibitem [{\citenamefont {{Hellings}}\ and\ \citenamefont
  {{Downs}}(1983)}]{1983ApJ...265L..39H}%
  \BibitemOpen
  \bibfield  {author} {\bibinfo {author} {\bibfnamefont {R.~W.}\ \bibnamefont
  {{Hellings}}}\ and\ \bibinfo {author} {\bibfnamefont {G.~S.}\ \bibnamefont
  {{Downs}}},\ }\bibfield  {title} {\bibinfo {title} {{Upper limits on the
  isotropic gravitational radiation background from pulsar timing analysis.}},\
  }\href {https://doi.org/10.1086/183954} {\bibfield  {journal} {\bibinfo
  {journal} {apjl}\ }\textbf {\bibinfo {volume} {265}},\ \bibinfo {pages} {L39}
  (\bibinfo {year} {1983})}\BibitemShut {NoStop}%
\bibitem [{\citenamefont {Bar-Kana}(1994)}]{PhysRevD.50.1157}%
  \BibitemOpen
  \bibfield  {author} {\bibinfo {author} {\bibfnamefont {R.}~\bibnamefont
  {Bar-Kana}},\ }\bibfield  {title} {\bibinfo {title} {Limits on direct
  detection of gravitational waves},\ }\href
  {https://doi.org/10.1103/PhysRevD.50.1157} {\bibfield  {journal} {\bibinfo
  {journal} {Phys. Rev. D}\ }\textbf {\bibinfo {volume} {50}},\ \bibinfo
  {pages} {1157} (\bibinfo {year} {1994})}\BibitemShut {NoStop}%
\bibitem [{\citenamefont {Gasperini}\ and\ \citenamefont
  {Veneziano}(1993)}]{GASPERINI1993317}%
  \BibitemOpen
  \bibfield  {author} {\bibinfo {author} {\bibfnamefont {M.}~\bibnamefont
  {Gasperini}}\ and\ \bibinfo {author} {\bibfnamefont {G.}~\bibnamefont
  {Veneziano}},\ }\bibfield  {title} {\bibinfo {title} {Pre-big-bang in string
  cosmology},\ }\href
  {https://doi.org/https://doi.org/10.1016/0927-6505(93)90017-8} {\bibfield
  {journal} {\bibinfo  {journal} {Astroparticle Physics}\ }\textbf {\bibinfo
  {volume} {1}},\ \bibinfo {pages} {317} (\bibinfo {year} {1993})}\BibitemShut
  {NoStop}%
\bibitem [{\citenamefont {Kuroyanagi}\ \emph {et~al.}(2012)\citenamefont
  {Kuroyanagi}, \citenamefont {Miyamoto}, \citenamefont {Sekiguchi},
  \citenamefont {Takahashi},\ and\ \citenamefont {Silk}}]{PhysRevD.86.023503}%
  \BibitemOpen
  \bibfield  {author} {\bibinfo {author} {\bibfnamefont {S.}~\bibnamefont
  {Kuroyanagi}}, \bibinfo {author} {\bibfnamefont {K.}~\bibnamefont
  {Miyamoto}}, \bibinfo {author} {\bibfnamefont {T.}~\bibnamefont {Sekiguchi}},
  \bibinfo {author} {\bibfnamefont {K.}~\bibnamefont {Takahashi}},\ and\
  \bibinfo {author} {\bibfnamefont {J.}~\bibnamefont {Silk}},\ }\bibfield
  {title} {\bibinfo {title} {Forecast constraints on cosmic string parameters
  from gravitational wave direct detection experiments},\ }\href
  {https://doi.org/10.1103/PhysRevD.86.023503} {\bibfield  {journal} {\bibinfo
  {journal} {Phys. Rev. D}\ }\textbf {\bibinfo {volume} {86}},\ \bibinfo
  {pages} {023503} (\bibinfo {year} {2012})}\BibitemShut {NoStop}%
\bibitem [{\citenamefont {Christensen}(2018)}]{Christensen_2019}%
  \BibitemOpen
  \bibfield  {author} {\bibinfo {author} {\bibfnamefont {N.}~\bibnamefont
  {Christensen}},\ }\bibfield  {title} {\bibinfo {title} {Stochastic
  gravitational wave backgrounds},\ }\href
  {https://doi.org/10.1088/1361-6633/aae6b5} {\bibfield  {journal} {\bibinfo
  {journal} {Reports on Progress in Physics}\ }\textbf {\bibinfo {volume}
  {82}},\ \bibinfo {pages} {016903} (\bibinfo {year} {2018})}\BibitemShut
  {NoStop}%
\bibitem [{\citenamefont {{FERMI-LAT Collaboration}}\ \emph
  {et~al.}(2022)\citenamefont {{FERMI-LAT Collaboration}} \emph
  {et~al.}}]{fermi2022gamma}%
  \BibitemOpen
  \bibfield  {author} {\bibinfo {author} {\bibnamefont {{FERMI-LAT
  Collaboration}}} \emph {et~al.},\ }\bibfield  {title} {\bibinfo {title} {{A
  gamma-ray pulsar timing array constrains the nanohertz gravitational wave
  background}},\ }\href {https://doi.org/10.1126/science.abm3231} {\bibfield
  {journal} {\bibinfo  {journal} {Science}\ }\textbf {\bibinfo {volume}
  {376}},\ \bibinfo {pages} {521} (\bibinfo {year} {2022})},\ \Eprint
  {https://arxiv.org/abs/2204.05226} {arXiv:2204.05226 [astro-ph.HE]}
  \BibitemShut {NoStop}%
\bibitem [{Note1()}]{Note1}%
  \BibitemOpen
  \bibinfo {note} {Https://zenodo.org/record/6374291\#.YzVcbC-KFpR}\BibitemShut
  {NoStop}%
\bibitem [{\citenamefont {{Aharonian}}\ \emph {et~al.}(2001)\citenamefont
  {{Aharonian}}, \citenamefont {{Konopelko}}, \citenamefont {{V{\"o}lk}},\ and\
  \citenamefont {{Quintana}}}]{aharonian2015}%
  \BibitemOpen
  \bibfield  {author} {\bibinfo {author} {\bibfnamefont {F.~A.}\ \bibnamefont
  {{Aharonian}}}, \bibinfo {author} {\bibfnamefont {A.~K.}\ \bibnamefont
  {{Konopelko}}}, \bibinfo {author} {\bibfnamefont {H.~J.}\ \bibnamefont
  {{V{\"o}lk}}},\ and\ \bibinfo {author} {\bibfnamefont {H.}~\bibnamefont
  {{Quintana}}},\ }\bibfield  {title} {\bibinfo {title} {{5@5 - a 5 GeV energy
  threshold array of imaging atmospheric Cherenkov telescopes at 5 km
  altitude}},\ }\href {https://doi.org/10.1016/S0927-6505(00)00164-X}
  {\bibfield  {journal} {\bibinfo  {journal} {Astroparticle Physics}\ }\textbf
  {\bibinfo {volume} {15}},\ \bibinfo {pages} {335} (\bibinfo {year} {2001})},\
  \Eprint {https://arxiv.org/abs/astro-ph/0006163} {arXiv:astro-ph/0006163
  [astro-ph]} \BibitemShut {NoStop}%
\bibitem [{\citenamefont {Aharonian}\ \emph {et~al.}(2021)\citenamefont
  {Aharonian}, \citenamefont {An}, \citenamefont {Axikegu} \emph
  {et~al.}}]{aharonian2021construction}%
  \BibitemOpen
  \bibfield  {author} {\bibinfo {author} {\bibfnamefont {F.}~\bibnamefont
  {Aharonian}}, \bibinfo {author} {\bibfnamefont {Q.}~\bibnamefont {An}},
  \bibinfo {author} {\bibnamefont {Axikegu}}, \emph {et~al.},\ }\bibfield
  {title} {\bibinfo {title} {Construction and on-site performance of the lhaaso
  wfcta camera},\ }\href@noop {} {\bibfield  {journal} {\bibinfo  {journal}
  {The European Physical Journal C}\ }\textbf {\bibinfo {volume} {81}},\
  \bibinfo {pages} {1} (\bibinfo {year} {2021})}\BibitemShut {NoStop}%
\bibitem [{\citenamefont {{Abdollahi}}\ \emph {et~al.}(2022)\citenamefont
  {{Abdollahi}}, \citenamefont {{Acero}}, \citenamefont {{Baldini}} \emph
  {et~al.}}]{abdollahi2022incremental}%
  \BibitemOpen
  \bibfield  {author} {\bibinfo {author} {\bibfnamefont {S.}~\bibnamefont
  {{Abdollahi}}}, \bibinfo {author} {\bibfnamefont {F.}~\bibnamefont
  {{Acero}}}, \bibinfo {author} {\bibfnamefont {L.}~\bibnamefont {{Baldini}}},
  \emph {et~al.},\ }\bibfield  {title} {\bibinfo {title} {{Incremental Fermi
  Large Area Telescope Fourth Source Catalog}},\ }\href
  {https://doi.org/10.3847/1538-4365/ac6751} {\bibfield  {journal} {\bibinfo
  {journal} {apjs}\ }\textbf {\bibinfo {volume} {260}},\ \bibinfo {eid} {53}
  (\bibinfo {year} {2022})},\ \Eprint {https://arxiv.org/abs/2201.11184}
  {arXiv:2201.11184 [astro-ph.HE]} \BibitemShut {NoStop}%
\bibitem [{\citenamefont {{Acero}}\ \emph {et~al.}(2016)\citenamefont
  {{Acero}}, \citenamefont {{Ackermann}}, \citenamefont {{Ajello}} \emph
  {et~al.}}]{acero2016development}%
  \BibitemOpen
  \bibfield  {author} {\bibinfo {author} {\bibfnamefont {F.}~\bibnamefont
  {{Acero}}}, \bibinfo {author} {\bibfnamefont {M.}~\bibnamefont
  {{Ackermann}}}, \bibinfo {author} {\bibfnamefont {M.}~\bibnamefont
  {{Ajello}}}, \emph {et~al.},\ }\bibfield  {title} {\bibinfo {title}
  {{Development of the Model of Galactic Interstellar Emission for Standard
  Point-source Analysis of Fermi Large Area Telescope Data}},\ }\href
  {https://doi.org/10.3847/0067-0049/223/2/26} {\bibfield  {journal} {\bibinfo
  {journal} {apjs}\ }\textbf {\bibinfo {volume} {223}},\ \bibinfo {eid} {26}
  (\bibinfo {year} {2016})},\ \Eprint {https://arxiv.org/abs/1602.07246}
  {arXiv:1602.07246 [astro-ph.HE]} \BibitemShut {NoStop}%
\bibitem [{\citenamefont {Sahakian}\ and\ \citenamefont
  {Akhperjanian}(2006)}]{cheback}%
  \BibitemOpen
  \bibfield  {author} {\bibinfo {author} {\bibfnamefont {V.}~\bibnamefont
  {Sahakian}}\ and\ \bibinfo {author} {\bibfnamefont {A.}~\bibnamefont
  {Akhperjanian}},\ }\bibfield  {title} {\bibinfo {title} {On the background of
  low threshold imaging cherenkov telescopes induced by cosmic ray electrons},\
  }\href {https://doi.org/10.1016/j.astropartphys.2006.06.009} {\bibfield
  {journal} {\bibinfo  {journal} {ASTROPARTICLE PHYSICS}\ }\textbf {\bibinfo
  {volume} {26}},\ \bibinfo {pages} {257} (\bibinfo {year} {2006})}\BibitemShut
  {NoStop}%
\bibitem [{\citenamefont {{Luo}}\ \emph {et~al.}(2021)\citenamefont {{Luo}},
  \citenamefont {{Ransom}}, \citenamefont {{Demorest}} \emph
  {et~al.}}]{luo2021pint}%
  \BibitemOpen
  \bibfield  {author} {\bibinfo {author} {\bibfnamefont {J.}~\bibnamefont
  {{Luo}}}, \bibinfo {author} {\bibfnamefont {S.}~\bibnamefont {{Ransom}}},
  \bibinfo {author} {\bibfnamefont {P.}~\bibnamefont {{Demorest}}}, \emph
  {et~al.},\ }\bibfield  {title} {\bibinfo {title} {{PINT: A Modern Software
  Package for Pulsar Timing}},\ }\href
  {https://doi.org/10.3847/1538-4357/abe62f} {\bibfield  {journal} {\bibinfo
  {journal} {apj}\ }\textbf {\bibinfo {volume} {911}},\ \bibinfo {eid} {45}
  (\bibinfo {year} {2021})},\ \Eprint {https://arxiv.org/abs/2012.00074}
  {arXiv:2012.00074 [astro-ph.IM]} \BibitemShut {NoStop}%
\bibitem [{\citenamefont {{Agazie}}\ \emph {et~al.}(2023)\citenamefont
  {{Agazie}}, \citenamefont {{Anumarlapudi}}, \citenamefont {{Archibald}} \emph
  {et~al.}}]{2023ApJ...951L...8A}%
  \BibitemOpen
  \bibfield  {author} {\bibinfo {author} {\bibfnamefont {G.}~\bibnamefont
  {{Agazie}}}, \bibinfo {author} {\bibfnamefont {A.}~\bibnamefont
  {{Anumarlapudi}}}, \bibinfo {author} {\bibfnamefont {A.~M.}\ \bibnamefont
  {{Archibald}}}, \emph {et~al.},\ }\bibfield  {title} {\bibinfo {title} {{The
  NANOGrav 15 yr Data Set: Evidence for a Gravitational-wave Background}},\
  }\href {https://doi.org/10.3847/2041-8213/acdac6} {\bibfield  {journal}
  {\bibinfo  {journal} {apjl}\ }\textbf {\bibinfo {volume} {951}},\ \bibinfo
  {eid} {L8} (\bibinfo {year} {2023})},\ \Eprint
  {https://arxiv.org/abs/2306.16213} {arXiv:2306.16213 [astro-ph.HE]}
  \BibitemShut {NoStop}%
\bibitem [{\citenamefont {{Janssen}}\ \emph {et~al.}(2015)\citenamefont
  {{Janssen}}, \citenamefont {{Hobbs}}, \citenamefont {{McLaughlin}} \emph
  {et~al.}}]{2015aska.confE..37J}%
  \BibitemOpen
  \bibfield  {author} {\bibinfo {author} {\bibfnamefont {G.}~\bibnamefont
  {{Janssen}}}, \bibinfo {author} {\bibfnamefont {G.}~\bibnamefont {{Hobbs}}},
  \bibinfo {author} {\bibfnamefont {M.}~\bibnamefont {{McLaughlin}}}, \emph
  {et~al.},\ }\bibfield  {title} {\bibinfo {title} {{Gravitational Wave
  Astronomy with the SKA}},\ }in\ \href {https://doi.org/10.22323/1.215.0037}
  {\emph {\bibinfo {booktitle} {Advancing Astrophysics with the Square
  Kilometre Array (AASKA14)}}}\ (\bibinfo {year} {2015})\ p.~\bibinfo {pages}
  {37},\ \Eprint {https://arxiv.org/abs/1501.00127} {arXiv:1501.00127
  [astro-ph.IM]} \BibitemShut {NoStop}%
\bibitem [{\citenamefont {{Pol}}\ \emph {et~al.}(2021)\citenamefont {{Pol}},
  \citenamefont {{Taylor}}, \citenamefont {{Kelley}} \emph
  {et~al.}}]{pol2021astrophysics}%
  \BibitemOpen
  \bibfield  {author} {\bibinfo {author} {\bibfnamefont {N.~S.}\ \bibnamefont
  {{Pol}}}, \bibinfo {author} {\bibfnamefont {S.~R.}\ \bibnamefont {{Taylor}}},
  \bibinfo {author} {\bibfnamefont {L.~Z.}\ \bibnamefont {{Kelley}}}, \emph
  {et~al.},\ }\bibfield  {title} {\bibinfo {title} {{Astrophysics Milestones
  for Pulsar Timing Array Gravitational-wave Detection}},\ }\href
  {https://doi.org/10.3847/2041-8213/abf2c9} {\bibfield  {journal} {\bibinfo
  {journal} {apjl}\ }\textbf {\bibinfo {volume} {911}},\ \bibinfo {eid} {L34}
  (\bibinfo {year} {2021})},\ \Eprint {https://arxiv.org/abs/2010.11950}
  {arXiv:2010.11950 [astro-ph.HE]} \BibitemShut {NoStop}%
\bibitem [{\citenamefont {{Sesana}}\ \emph {et~al.}(2004)\citenamefont
  {{Sesana}}, \citenamefont {{Haardt}}, \citenamefont {{Madau}},\ and\
  \citenamefont {{Volonteri}}}]{2004ApJ...611..623S}%
  \BibitemOpen
  \bibfield  {author} {\bibinfo {author} {\bibfnamefont {A.}~\bibnamefont
  {{Sesana}}}, \bibinfo {author} {\bibfnamefont {F.}~\bibnamefont {{Haardt}}},
  \bibinfo {author} {\bibfnamefont {P.}~\bibnamefont {{Madau}}},\ and\ \bibinfo
  {author} {\bibfnamefont {M.}~\bibnamefont {{Volonteri}}},\ }\bibfield
  {title} {\bibinfo {title} {{Low-Frequency Gravitational Radiation from
  Coalescing Massive Black Hole Binaries in Hierarchical Cosmologies}},\ }\href
  {https://doi.org/10.1086/422185} {\bibfield  {journal} {\bibinfo  {journal}
  {apj}\ }\textbf {\bibinfo {volume} {611}},\ \bibinfo {pages} {623} (\bibinfo
  {year} {2004})},\ \Eprint {https://arxiv.org/abs/astro-ph/0401543}
  {arXiv:astro-ph/0401543 [astro-ph]} \BibitemShut {NoStop}%
\bibitem [{\citenamefont {{Fan}}\ \emph {et~al.}(2022)\citenamefont {{Fan}},
  \citenamefont {{Chang}}, \citenamefont {{Guo}} \emph
  {et~al.}}]{2022AcASn..63...27F}%
  \BibitemOpen
  \bibfield  {author} {\bibinfo {author} {\bibfnamefont {Y.~Z.}\ \bibnamefont
  {{Fan}}}, \bibinfo {author} {\bibfnamefont {J.}~\bibnamefont {{Chang}}},
  \bibinfo {author} {\bibfnamefont {J.~H.}\ \bibnamefont {{Guo}}}, \emph
  {et~al.},\ }\bibfield  {title} {\bibinfo {title} {{Very Large Area Gamma-ray
  Space Telescope (VLAST)}},\ }\href {https://doi.org/2022AcASn..63...27F}
  {\bibfield  {journal} {\bibinfo  {journal} {Acta Astronomica Sinica}\
  }\textbf {\bibinfo {volume} {63}},\ \bibinfo {eid} {27} (\bibinfo {year}
  {2022})}\BibitemShut {NoStop}%
\bibitem [{\citenamefont {Hofmann}\ and\ \citenamefont
  {Zanin}(2023)}]{hofmann2023cherenkov}%
  \BibitemOpen
  \bibfield  {author} {\bibinfo {author} {\bibfnamefont {W.}~\bibnamefont
  {Hofmann}}\ and\ \bibinfo {author} {\bibfnamefont {R.}~\bibnamefont
  {Zanin}},\ }\href@noop {} {\bibinfo {title} {The cherenkov telescope array}}
  (\bibinfo {year} {2023}),\ \Eprint {https://arxiv.org/abs/2305.12888}
  {arXiv:2305.12888 [astro-ph.IM]} \BibitemShut {NoStop}%
\bibitem [{\citenamefont {Zhen}\ \emph {et~al.}(2019)\citenamefont {Zhen},
  \citenamefont {Ming-Jun}, \citenamefont {Song-Zhan}, \citenamefont {Hong-Bo},
  \citenamefont {Cheng}, \citenamefont {Ye}, \citenamefont {Ling-Ling},
  \citenamefont {Xin-Hua}, \citenamefont {Xiang-Dong}, \citenamefont {Han-Rong}
  \emph {et~al.}}]{zhen2019introduction}%
  \BibitemOpen
  \bibfield  {author} {\bibinfo {author} {\bibfnamefont {C.}~\bibnamefont
  {Zhen}}, \bibinfo {author} {\bibfnamefont {C.}~\bibnamefont {Ming-Jun}},
  \bibinfo {author} {\bibfnamefont {C.}~\bibnamefont {Song-Zhan}}, \bibinfo
  {author} {\bibfnamefont {H.}~\bibnamefont {Hong-Bo}}, \bibinfo {author}
  {\bibfnamefont {L.}~\bibnamefont {Cheng}}, \bibinfo {author} {\bibfnamefont
  {L.}~\bibnamefont {Ye}}, \bibinfo {author} {\bibfnamefont {M.}~\bibnamefont
  {Ling-Ling}}, \bibinfo {author} {\bibfnamefont {M.}~\bibnamefont {Xin-Hua}},
  \bibinfo {author} {\bibfnamefont {S.}~\bibnamefont {Xiang-Dong}}, \bibinfo
  {author} {\bibfnamefont {W.}~\bibnamefont {Han-Rong}}, \emph {et~al.},\
  }\bibfield  {title} {\bibinfo {title} {Introduction to large high altitude
  air shower observatory (lhaaso)},\ }\href@noop {} {\bibfield  {journal}
  {\bibinfo  {journal} {Chinese Astronomy and Astrophysics}\ }\textbf {\bibinfo
  {volume} {43}},\ \bibinfo {pages} {457} (\bibinfo {year} {2019})}\BibitemShut
  {NoStop}%
\bibitem [{\citenamefont {Li}\ \emph {et~al.}(2023)\citenamefont {Li},
  \citenamefont {Feng}, \citenamefont {Zhang} \emph {et~al.}}]{lacticrc}%
  \BibitemOpen
  \bibfield  {author} {\bibinfo {author} {\bibfnamefont {K.}~\bibnamefont
  {Li}}, \bibinfo {author} {\bibfnamefont {S.}~\bibnamefont {Feng}}, \bibinfo
  {author} {\bibfnamefont {S.}~\bibnamefont {Zhang}}, \emph {et~al.},\
  }\bibfield  {title} {\bibinfo {title} {Development of aluminum honeycomb
  reflector in lact},\ }\href@noop {} {\bibfield  {journal} {\bibinfo
  {journal} {Proc. Sci., ICRC2023}\ }\textbf {\bibinfo {volume} {219}}
  (\bibinfo {year} {2023})}\BibitemShut {NoStop}%
\bibitem [{\citenamefont {Deng}\ \emph {et~al.}(2021)\citenamefont {Deng},
  \citenamefont {Yang}, \citenamefont {Chen} \emph {et~al.}}]{deng2021lenghu}%
  \BibitemOpen
  \bibfield  {author} {\bibinfo {author} {\bibfnamefont {L.}~\bibnamefont
  {Deng}}, \bibinfo {author} {\bibfnamefont {F.}~\bibnamefont {Yang}}, \bibinfo
  {author} {\bibfnamefont {X.}~\bibnamefont {Chen}}, \emph {et~al.},\
  }\bibfield  {title} {\bibinfo {title} {Lenghu on the tibetan plateau as an
  astronomical observing site},\ }\href@noop {} {\bibfield  {journal} {\bibinfo
   {journal} {Nature}\ }\textbf {\bibinfo {volume} {596}},\ \bibinfo {pages}
  {353} (\bibinfo {year} {2021})}\BibitemShut {NoStop}%
\end{thebibliography}%

\end{document}